\documentclass[twocolumn,showpacs,preprintnumbers,amsmath,amssymb]{revtex4}

\usepackage{graphicx}
\usepackage[dvips]{color}

\begin{document}


\title{A simple mean field equation for condensates in the BEC-BCS crossover regime}

\author{Cheng Chin}

\affiliation{Institut f\"{u}r Experimentalphysik, Universit\"{a}t
Innsbruck, Technikerstr. 25, 6020 Innsbruck, Austria
}

\date{\today}

\begin{abstract}
We present a mean field approach based on pairs of fermionic atoms
to describe condensates in the BEC-BCS crossover regime. By
introducing an effective potential, the mean field equation allows
us to calculate the chemical potential, the equation of states and
the atomic correlation function. The results agree surprisingly
well with recent quantum Monte Carlo calculations. We show that
the smooth crossover from the bosonic mean field repulsion between
molecules to the Fermi pressure among atoms is associated with the
evolution of the atomic correlation function.
\end{abstract}



\maketitle

Recent studies on ultracold Fermi gases and molecular condensates
\cite{mbec} address an intriguing topic, the crossover from a
Bose-Einstein condensate of composite bosons to a fermionic
Bardeen-Cooper-Schrieffer superfluid (BEC-BCS crossover)
\cite{becbcs}. By magnetically tuning the interaction strength
near a Feshbach resonance \cite{feshbach}, a molecular BEC can be
smoothly converted into a degenerate Fermi gas and vice versa.
Experimental \cite{xoverexpinn, xoverexp1, xoverexp2} and
theoretical research \cite{xovertheo, resfluid} into the quantum
gases in the crossover regime are highly active and may provide
new insights into other strongly interacting Fermi systems.

In contrast to weakly interacting atomic BECs, for which a simple
mean field description based on the Gross-Pitaevskii equation has
been very successful \cite{stringari}, theoretical models on the
condensates in the crossover regime are generally very
sophisticated and require expertise borrowed from condensed matter
theory. The difficulty in providing a simple model for the
fermionic system comes from, firstly, the lack of a small
expansion parameter. The full range of atomic scattering length
$a$ should be taken into account to describe the crossover.
Secondly, quantum many-body correlations are intrinsically more
complicated for fermionic systems than for bosonic ones.

The BEC-BCS crossover, however, suggests an alternative approach
to model the strongly interacting fermions based on composite
bosons. This is possible since a Fermi gas in the crossover regime
constitutes the same quantum phase as of a condensate of
interacting pairs. Recent experiments on the wave function
projection \cite{xoverexp1} and on the pairing gap \cite{gap}
indeed indicate that near the Feshbach resonance, a large fraction
of fermionic atoms are paired at low temperatures. From these
observations, we propose a bosonic mean field equation,
complementary to the fermion-based BCS approaches, to describe the
atom pairs in the crossover regime. Our mean-field approach is
relatively simple and well-behaved near the resonance. We obtain
analytic expressions for the chemical potential and the equation
of states, which agree very well with other calculations. In
particular, we find the chemical potential in the unitarity limit
is $\sim0.4357$ times that in the BCS limit, in excellent
agreement with the recent quantum Monte Carlo calculations of
$0.42\sim0.44$ \cite{carlson, giorgini}.

We consider an ultracold gas of two-component fermionic atoms. At
low temperatures, only atoms in different internal states can pair
via $s$-wave interaction. For simplicity, we assume the
interaction range is zero. In the absence of many-body effects,
the center-of-mass motion of an atom pair $\Psi_0(\vec{R})$ is
decoupled from the internal relative atomic motion
$\psi_0(\vec{r})=(4\pi r^2)^{-1/2}\psi_0(r)$ with $r=|\vec{r}|$
the atomic separation. Given the atomic scattering length $a$,
$\psi_0(r)$ satisfies Schr\"{o}dinger's equation,

\begin{equation}
-\frac{\hbar^2}{m}\psi_0''(r) = -E_b\psi_0(r)
\end{equation}
with the boundary condition $\psi_0(0)=-a\psi_0'(0)$. Here $m$ is
the atomic mass, $2\pi\hbar$ is Planck's constant, and $E_b$ is
the molecular binding energy.

For positive scattering lengths $a>0$, the bound state is
described by $\psi_0=(2/a)^{1/2}e^{-r/a}$ with
$E_b=\hbar^2/(ma^2)$. The size of the molecule is given by
$\langle r\rangle=a/2$. For negative scattering lengths $a<0$, the
bound state does not exist and the ground state energy is
$-E_b=0$.

Now consider a condensate of pairs with a density distribution
$n(\vec{R})$ in a slow-varying potential well $V(\vec{R})$. We
introduce the many-body wave function to include the condensate of
the bosonic pairs $\Psi(\vec{R})=n(\vec{R})^{1/2}$ as well as the
internal atomic correlation $\psi(r)$. The mean field equation for
the composite bosons is then

\begin{eqnarray}
&(-\frac{\hbar^2\nabla_R^2}{4m}-\frac{\hbar^2\partial_r^2}{m}+V+\hat{U}\,)\Psi(\vec{R})\psi(r)=\mu_m\Psi(\vec{R})\psi(r),\\
&\psi(0)=-a\,\partial_r\psi(0).
\end{eqnarray}
Here $\mu_m$ is the chemical potential,
$\hat{U}=\hat{g}|\Psi(\vec{R})|^2$ is the mean field interaction
and $\hat{g}$ is the interaction term.

In conventional approaches, $\hat{g}$ is given by the scattering
length of the bosons. For pairs of fermions, scattering length
$a_m$ is determined by that of the constituent atoms as
$a_m=0.60a$, resulting from an effective repulsive potential
between molecules \cite{petrov}. This dependence can be understood
in a simple picture. Low-energy collision with a repulsive
interactions acquires a scattering length which is proportional to
the size of the scatterer. For pairs of atoms, we have
$a_m\sim\langle r\rangle=a/2$.

From the above considerations, we hypothesize that the interaction
term $\hat{g}$ is effectively proportional to the interatomic
separation $r$ as

\begin{equation}
\hat{g}=g(r)=c \frac{\hbar^2}{m} r,
\end{equation}
where $c$ is a dimensionless constant.

To proceed with minimum algebra, we consider a uniform gas with a
density $|\Psi(\vec{R})|^2=n$. Eq.~(2) becomes

\begin{equation}
(-\frac{\hbar^2}{m}\partial_r^2+\hat{g}\,)\psi(r)=\mu_m\psi(r).
\end{equation}

To determine $c$, we consider the BEC limit ($na_m^3\ll 1$), where
the mean field term can be treated perturbatively. That is, the
expectation value of $\hat{U}$ based on the bare molecular wave
function $\psi_0(r)=(2/a)^{1/2}e^{-r/a}$ should yield the
molecular mean field shift $4\pi\hbar^2a_mn/2m$,

\begin{equation}
\int_0^{\infty} \psi_0^*(r) \hat{g}\,n \psi_0(r)
dr=\frac{2\pi\hbar^2a_mn}{m}.
\end{equation}

Using $a_m=0.60a$, we find Eq.~(6) can indeed hold for arbitrary
scattering lengths $a_m$ based on the linear mean field potential
in Eq.~(4). We determine $c=4\pi a_m/a\approx 7.5$.


From Eq.(3), (4), and (5), the exact solution of the pair wave
function is given by


\begin{eqnarray}
\psi(r)&=&N \mbox{Ai}(c^{1/3}n^{1/3}r-c^{-2/3}\mu_m/E_0) \\
\psi(0)&=&-a~\partial_r\psi(0),
\end{eqnarray}
where $N$ is the normalization constant, Ai$(x)$ is Airy's Ai
function, and $E_0=\hbar^2n^{2/3}/m$. Notice that the chemical
potential $\mu_m$ in Eq.~(7) is determined from Eq.~(8).

In the weak interaction limit $0 < na_m^3\ll 1$, the wave function
$\psi(r)$ obtained from Eq.(7) is identical to the unperturbed one
$\psi_0(r)$ for $r\ll n^{-1/3}$. For $r\gg
 n^{-1/3}$, $\psi(r)$ is exponentially smaller than $\psi_0(r)$ and
approaches $\sim r^{-1/4}$exp$(-\frac23r^{3/2})$. This suppression
for large atomic separation is expected since the interaction
energy increases when the pairs start overlapping. As a
consequence, the pair wave function $\psi(r)$ is compressed to a
smaller size than that of a bare molecule. Similar effect is also
discussed in Ref.~\cite{squeeze}

The distortion of the pair wave function can be characterized by
an effective shift in the binding energy $E_b$. In the weak
interaction limit, the shift can be defined as

\begin{equation}
\int_0^{\infty} \psi^*(r) (-\frac{\hbar^2\partial_r^2}{m})\psi(r)
dr= -E_b(1+o(na^3)).
\end{equation}
The binding energy correction $o(na^3)$ is positive. This effect
is absent in the calculations for point-like bosons \cite{bosongp}
since it originates comes from the internal degree of freedom.
This increase in binding energy is expected since the pair is
compressed. This result also provides a simple picture to
understand the augmentation of the molecular binding energy
reported in Ref.~\cite{gap}.

We extend the mean field model to the crossover and the BCS
regime, where the atom pairs strongly overlap. Although it becomes
less clear if the mean field approach can fully capture the
Fermionic nature of the gas, our aim here is to determine an
effective potential which can best describe the system in the
BEC-BCS crossover regime.

In this regime, the four-body calculation of $a_m=0.6a$ is no
longer valid, and we determine the mean field interaction
$\hat{U}$ from the properties of the Fermi gas. First of all, in
the dilute gas limit, we still expect the interaction to be
proportional to the square of the bosonic field,
$\hat{U}=\hat{g}|\Psi(R)|^2$. Secondly, we exploit the asymptotic
behavior of the gas in the weak coupling limit $na^3\rightarrow
0^-$, where the system approaches an ideal degenerate Fermi gas
with the chemical potential

\begin{equation}
\lim_{na^3\rightarrow0^-}
\mu_m=2E_\mathrm{F}=\frac{(6\pi^2n)^{2/3}\hbar^2}{m},
\end{equation}
where $E_\mathrm{F}=\hbar^2k_\mathrm{F}^2/2m$ is the Fermi energy
and $k_\mathrm{F}=(6\pi^2n)^{1/3}$ is the Fermi wave number.

Based on Eq.~(5), we find that the above density dependence
$\mu_m\propto n^{2/3}$ can be satisfied only when the interaction
term $\hat{g}$ is again a linear function of $r$. Taking the limit
of $a=0^-$ and assuming $g(r)=c'(\hbar^2/m) r$, we can solve the
chemical potential from Eq.(7) and Eq.(8) as $\mu_m=\alpha
c'^{2/3}E_0$, where $-\alpha\approx-2.338$ is the first zero of
the Ai$(x)$ function. Equating $\mu_m$ to $2E_\mathrm{F}$ yields
$c'=6\pi^2\alpha^{-3/2}\approx16.56$. This value is about twice as
large as $c$.

We first test the equation in the unitarity limit $a=\pm\infty$.
Fermi gases in this limit have been extensively studied, for which
a universal and fermionic behavior is expected \cite{resfluid}.
Due to the divergence of the scattering length, we expect the only
energy scale in the system is the Fermi energy $E_\mathrm{F}$.
From the boundary condition $\partial_r\psi(0)=0$ and
$g=c'(\hbar^2/m) r$, we determine the chemical potential as
$\mu_m=\alpha'c'^{2/3}E_0$, where $-\alpha'\approx-1.019$ is the
first zero of the Ai$'(x)$ function. Given
$c'=6\pi^2\alpha^{-3/2}$, we get $\mu_m/2=(\alpha'/\alpha)
E_\mathrm{F}\approx 0.4357E_\mathrm{F}$. This result agrees
excellently with recent quantum Monte Carlo calculations which
gives $\mu_m/2=0.44(1) E_\mathrm{F}$\cite{carlson} and
$0.42(1)E_\mathrm{F}$ \cite{giorgini}, and the measurements
\cite{xoverexpinn, betaens}, where the uncertainties are larger.
We, however, cannot exclude this agreement is coincidental. Near
the unitarity limit, we have

\begin{eqnarray}
\frac{\mu_m}{2E_\mathrm{F}}=\frac{\alpha'}{\alpha}-\frac{\alpha^{-1/2}}{\alpha'k_\mathrm{F}a}+O(\frac{1}{k_\mathrm{F}^2a^2}).
\end{eqnarray}

Next, we investigate the BEC-BCS crossover regime. Rewritting
Eq.~(7) and (8) using $c'=6\pi^2\alpha^{-3/2}$, we get

\begin{eqnarray}
\psi(r)&=&N\,\mbox{Ai}(\alpha^{-1/2}k_\mathrm{F}r-\alpha\frac{\mu_m}{2E_\mathrm{F}}) \\
\frac{k_\mathrm{F}a}{\alpha^{1/2}}&=&-\frac{\mbox{Ai}(-\alpha\mu_m/2E_\mathrm{F})}{\mbox{Ai}'(-\alpha\mu_m/2E_\mathrm{F})}.
\end{eqnarray}

\begin{figure}
\includegraphics[width=2.8in]{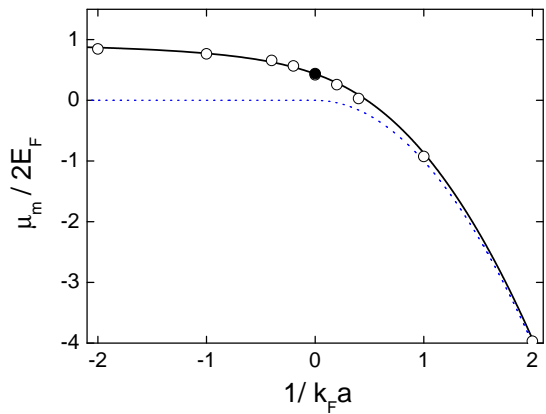}
\caption{Chemical potential $\mu_m$ in the crossover regime (solid
line). For large $1/k_\mathrm{F}a$, the chemical potential $\mu_m$
approaches the energy of the molecular state $-E_b$ (dotted line).
The solid dots and the open dots show the Monte Carlo calculations
from Ref.~\cite{carlson} and Ref.~\cite{giorgini}, respectively.}
\label{fig1}
\end{figure}

The chemical potential $\mu_m$ calculated from Eq.(13) is shown in
Fig.~\ref{fig1}. We see that $\mu_m$ approaches $2E_\mathrm{F}$ in
the BCS limit and $-E_b$ in the BEC limit, as expected. In the
crossover regime, the values agree well with the Monte Carlo
calculation from \cite{giorgini}. We can also evaluate the
equation of states $\mu_m+E_b\propto n^{\gamma}$, where the
exponent $\gamma$ plays a crucial role in the collective
excitation frequencies \cite{stringari, heiselberg}. From Eq.(12)
and Eq.(13), we obtain

\begin{eqnarray}
\gamma&=&\frac{d\ln(\mu_m+E_b)}{d\ln n}\\
      &=&\frac23\left( 1+\frac{E_b}{\mu_m}\right)^{-1}\left(1+\frac{2\alpha^{-3/2}
k_\mathrm{F}aE_\mathrm{F}^2/\mu_m^2}{k_\mathrm{F}^2a^2+2E_\mathrm{F}/\mu_m}\right).
\end{eqnarray}


\begin{figure}
\includegraphics[width=2.8in]{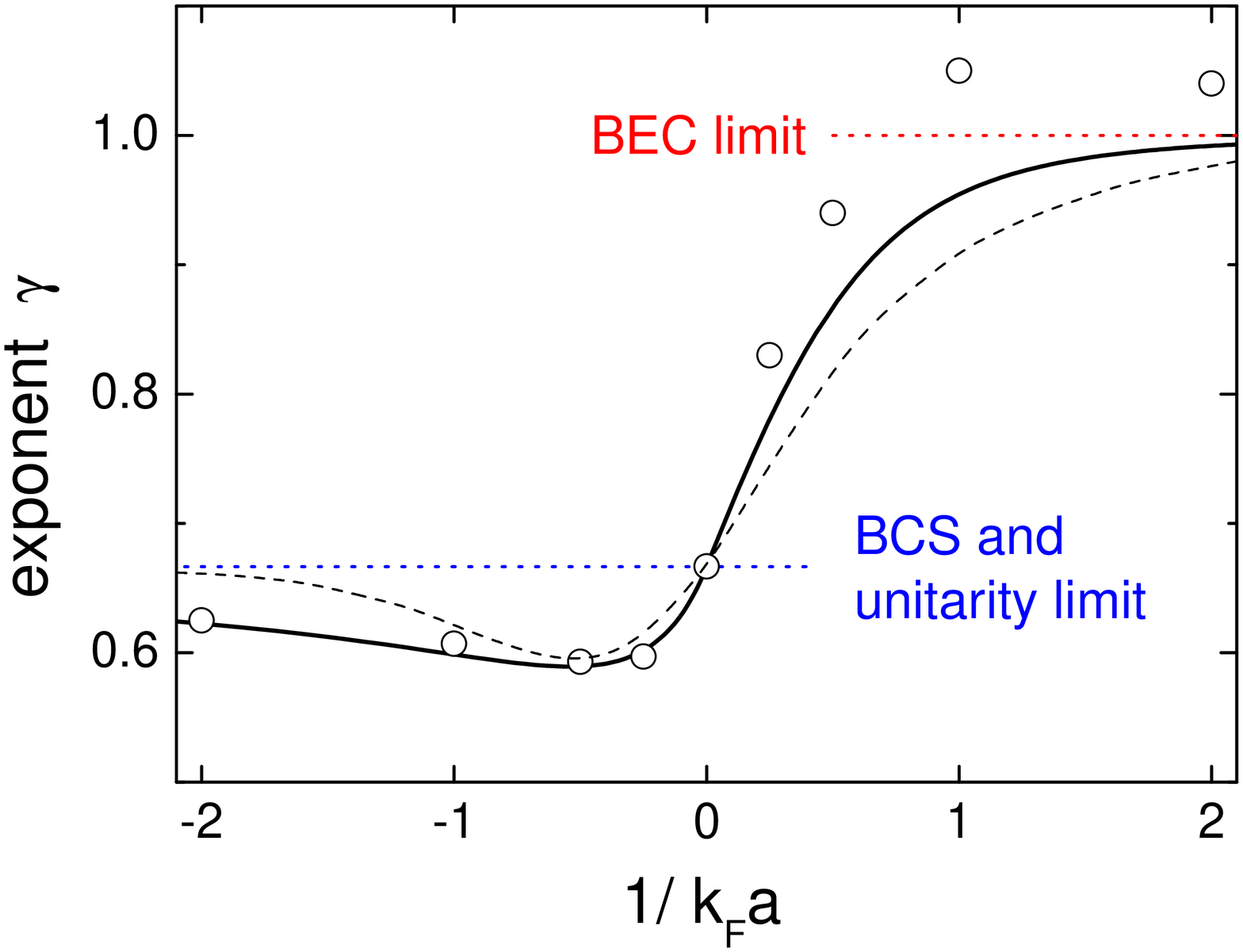}
\caption{Exponent $\gamma$ for the equation of states from the
mean field calculation (solid line), the BCS calculation (dashed
line) \cite{hui} and the fit to the quantum Monte Carlo
calculation (open circles) \cite{giorgini, manini}. The unitarity
and BCS limit $\gamma=2/3$ and the BEC limit $\gamma=1$ are shown
in dotted lines. } \label{fig1.5}
\end{figure}

\begin{figure}
\includegraphics[width=2.8in]{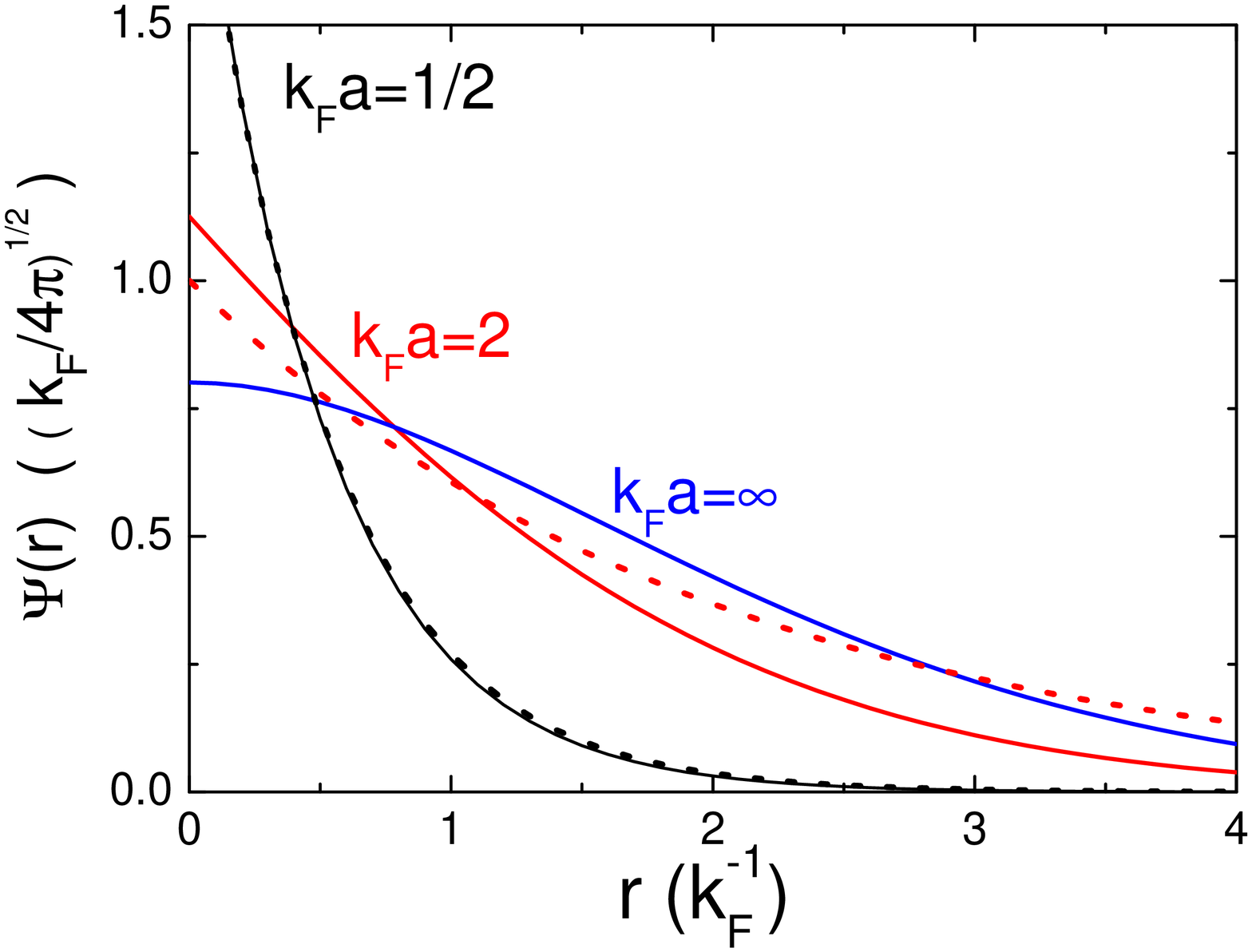}
\caption{Pair wave functions in the crossover regime. Wave
functions $\psi(r)$ at $k_\mathrm{F}a=1/2$, $k_\mathrm{F}a=2$ and
$k_\mathrm{F}a=\pm\infty$, shown in solid lines, are calculated
based on Eq.(12). In the former two cases, the bare molecular wave
functions $\psi_0(r)$ (dotted lines) are shown for comparison.}
\label{fig2}
\end{figure}

The exponent $\gamma$ (see Fig.~\ref{fig1.5}) shows the expected
behavior: $\gamma=1$ in the BEC limit and $\gamma=2/3$ in the BCS
and unitarity limits. In the range of $1<k_\mathrm{F}a<\infty$,
$\gamma$ shows a dramatic variation. In the following, we show
that this dramatic variation is directly linked to the crossover
nature of the quantum gas and is a result of the distortion of the
pair wave function $\psi(r)$.

From Eq.~(12), we calculate $\psi(r)$ for $k_\mathrm{F}a=1/2$, 2,
and $\pm\infty$, shown in Fig.~\ref{fig2}. For
$k_\mathrm{F}a=1/2$, we see very small deviation of $\psi(r)$ from
the bare molecular wave function $\psi_0(r)$. For
$k_\mathrm{F}a=2$, $\psi(r)$ is clearly different from $\psi_0(r)$
with a higher probability amplitude for $r<k_\mathrm{F}^{-1}$ and
a lower amplitude for $r>k_\mathrm{F}^{-1}$. This is the
compression effect we discussed. In the unitarity limit
$k_\mathrm{F}a=\pm\infty$, the atomic pairing is fermionic since
bare molecules dissociate at this point. The mean atomic
separation of $\langle
r\rangle\approx2/k_\mathrm{F}\approx0.5n^{-1/3}$ suggests the size
of the pairs is about half of the mean molecular spacing.

The distortion of the wave functions leads to significant
consequences for the quantum gas. Given the mean field energy as
$\langle \hat{U}\rangle \propto n\langle r\rangle$, the evolution
of the pair size from $\langle r\rangle \propto a$ in the BEC
regime to $\langle r\rangle \propto n^{-1/3}$ in the unitarity
limit underlies the crossover nature of the interactions from the
bosonic mean field repulsion $\langle \hat{U}\rangle \propto na$
to the Fermi pressure $\langle \hat{U}\rangle \propto n^{2/3}$.
This explains the variation of the exponent $\gamma$ in Fig.~(2).
From these observations, we can qualitatively define the BCS
regime to be $k_\mathrm{F}a<0$, crossover regime
$1/2<k_\mathrm{F}a<\infty$, and BEC regime $0<k_\mathrm{F}a<1/2$.
The use of $c'\approx16.56$ in the mean field term is appropriate
in the BCS and crossover regimes and $c\approx7.5$ in the BEC
regime.


The pair wave functions can be directly probed experimentally by
radio-frequency (rf) excitations as demonstrated in Ref.
\cite{jinrf, gap}. In these experiments, rf photons excite the
bound pairs into another spin state in which no bound state
exists. The excited pairs then dissociate into free atoms.
Theoretical calculation based on bare molecules show that the
excitation rate constant, or the bound-free Franck-Condon factor
$F_f(K)$, reflects the pair wave function in the momentum space
\cite{rftheory},

\begin{eqnarray}
 F_f(K)&=& \frac{m}{\pi\hbar^2 k}\left | \int_0^{\infty} \sin{(kr+\delta)} \psi_0(r) dr \right |^2,
\end{eqnarray}
where $K=\hbar^2k^2/m$, $k$, and $\delta$ are the energy, relative
wave number and the scattering phase shift of the outgoing atoms,
respectively.

\begin{figure}
\includegraphics[width=2.6in]{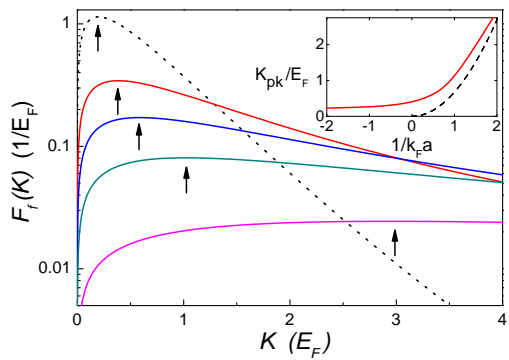}
\caption{Bound-free Franck-Condon factors $F_f(K)$ of the pairs
for (from bottom to top) $k_\mathrm{F}a=1/2$, $k_\mathrm{F}a=1$,
$k_\mathrm{F}a=2$, $k_\mathrm{F}a=\pm\infty$ (unitarity limit) and
$k_\mathrm{F}a=0^-$ (BCS limit, dotted line). The arrows mark the
peak positions $K_{pk}$. In the inset, $K_{pk}$ is plotted as a
function of $1/k_\mathrm{F} a$ (solid line) together with the
$K_{pk}$ for bare molecules (dashed line).} \label{fig3}
\end{figure}

To calculate Franck-Condon factors in the crossover regime, we
replace $\psi_0(r)$ by $\psi(r)$ and assume the atoms in the
outgoing channel do not interact $\delta=0$. In Fig.~\ref{fig3},
we show that the Franck-Condon factors display a resonance
structure in the crossover regime. The location of the peak
Franck-Condon factor $K_{pk}$ provides a sensitive measure of the
atomic correlation length. In the BEC regime, $K_{pk}$ approaches
$\frac43E_b\gg E_\mathrm{F}$ \cite{rftheory} and suggests that the
atomic separation is small compared to the intermolecular
distance. In the crossover regime, $K_{pk}$ approaches a small
fraction of $E_\mathrm{F}$. The persistence of the resonance
structure at unitarity and in the BCS regime indicates the
correlation of the atoms in momentum space. This dependence is
recently reported in \cite{gap, jingap}. A quantitative comparison
with the measurements, however, must include the effects of the
trapping potential and the finite temperature \cite{paivi}, which
is outside the scope of this paper.

In conclusion, we provide a simple mean field model to describe
the BEC-BCS crossover. From this equation, many properties of the
strongly interacting gas can be analytically calculated with high
accuracy. Our model can also be easily generalized to include the
external potential and to study crossover effects in systems with
lower or higher dimension.


We thank G.V. Shlyapnikov for stimulating discussions and the
members in R. Grimm's Li group in Innsbruck for their support.
C.C. is a Lise-Meitner research fellow of the Austrian Science
Fund (FWF).


\begin{references}
\bibitem{mbec}
S. Jochim{\it et al.},
Science {\bf 302}, 2101 (2003); published online November 13, 2003
(10.1126/science.1093280); M. Greiner, C. A. Regal, D. S. Jin,
Nature {\bf 426}, 537 (2003); Zwierlein{\it et al.},
Phys. Rev. Lett. {\bf 91}, 250401 (2003).

\bibitem{becbcs}
D. M. Eagles, Phys. Rev. {\bf 186}, 456 (1969); A. J. Leggett, in
{\it Modern Trends in the Theory of Condensed Matter},
(Springer-Verlag, Berlin, 1980); P. Nozi\`eres, S. Schmitt-Rink,
{J. Low Temp. Phys.} {\bf 59}, 195 (1985); Q. Chen, J. Stajic, S.
Tan and K. Levin, cond-mat/0404274.

\bibitem{feshbach}
E. Tiesinga, B. J. Verhaar, and H. T. C. Stoof, Phys. Rev. A {\bf
47}, 4114 (1993).

\bibitem{xoverexpinn}
M. Bartenstein{\it et al.},
Phys. Rev. Lett. {\bf 92}, 120401 (2004).

\bibitem{xoverexp1}
C. A. Regal, M. Greiner, D. S. Jin, Phys. Rev. Lett. {\bf 92},
040403 (2004); M. W. Zwierlein{\it et al.},
Phys. Rev. Lett. {\bf 92}, 120403 (2004)
\bibitem{xoverexp2}
J. Kinast{\it et al.},
Phys. Rev. Lett. {\bf 92}, 150402 (2004); M. Bartenstein{\it et
al.},
Phys. Rev. Lett. {\bf 92}, 203201 (2004).

\bibitem{xovertheo}
E. Timmermans, K. Furuya, P. W. Milonni, A. K. Kerman, Phys. Lett.
A {\bf 285}, 228 (2001); H. Heiselberg, Phys. Rev. A {\bf 63},
043606 (2001); Y. Ohashi, A. Griffin, Phys. Rev. Lett. {\bf 89},
130402 (2002); A. Perali, P. Pieri, and G.C. Strinati, Phys. Rev.
A {\bf 68}, 031601(R) (2003); J. Stajic {\it et al.}, Phys. Rev. A
{\bf 69}, 063610 (2004).

\bibitem{resfluid}
H. Heiselberg, Phys. Rev. A {\bf 63}, 043606 (2001); M. Holland,
S.J.J.M.F. Kokkelmans, M. L. Chiofalo, R. Walser, Phys. Rev. Lett.
{\bf 87}, 120406 (2001); T.-L. Ho, Phys. Rev. Lett. {\bf 92},
090402 (2004).

\bibitem{stringari}
L. Pitaevski and S. Stringari, {\it Bose-Einstein Condensation},
(Clarendon, Oxford, 2003).

\bibitem{gap}
C. Chin{\it et al.},
Science {\bf 305}, 1128 (2004); published online July 22, 2004;
(10.1126/science.1100818)

\bibitem{jingap}
M. Greiner, C.A. Regal and D.S. Jin,
cond-mat/0407381.

\bibitem{carlson} J. Carlson, S.-Y. Chang, V.R.
Pandharipande, and K.E. Schmidt, Phys. Rev. Lett. {\bf 91}, 050401
(2003).

\bibitem{giorgini} G.E. Astrakharchik, J. Boronat, J. Casulleras, and S. Giorgini,
cond-mat/0406113.

\bibitem{petrov}
D.S. Petrov, C. Salomon and G.V. Shlyapnikov, Phys. Rev. Lett.
{\bf 93}, 090404 (2004); D.S. Petrov, C. Salomon and G.V.
Shlyapnikov, cond-mat/0407579

\bibitem{squeeze}
N. Andrenacci, P. Pieri, and G.C. Strinati, Eur. Phys. Jour. B
{\bf 13}, 637 (2000).

\bibitem{bosongp}
A. Fetter and J.D. Walecka, \emph{Quantum Theory of Many-Particle
Systems}, (Dover, New York, 2003).

\bibitem{betaens}
K.M. O'Hara et al., Science {\bf 298}, 2179 (2002); M.E. Gehm et
al., Phys. Rev. A 68, 011401 (2003); T. Bourdel{\it et al.},
Phys. Rev. Lett. {\bf 93}, 050401 (2004).
\bibitem{heiselberg} H. Heiselberg, Phys. Rev. Lett. {\bf 93}, 040402 (2004)

\bibitem{hui} H. Hu et al., Phys. Rev. Lett. {\bf 93}, 190403 (2004).

\bibitem{manini} N. Manini, L. Salasnich, cond-mat/0407039.



\bibitem{rftheory}
C. Chin and P. Julienne, cond-mat/0408254.

\bibitem{jinrf}
C.A. Regal, C. Ticknor, J.L. Bohn, D.S. Jin, Nature {\bf 424}, 47
(2003).

\bibitem{paivi}
J. Kinnunen, M. Rodríguez, and P. T\"{o}rm\"{a}, Science {\bf
305}, 1131 (2004); published online July 22, 2004;
(10.1126/science.1100782)




\end{references}
\end{document}